\documentstyle[twocolumn,aps,epsfig,floats]{revtex}
\draft
\begin{document}
%=======================================================================
\twocolumn[\hsize\textwidth\columnwidth\hsize\csname @twocolumnfalse\endcsname

\title{Resonant photoemission spectroscopy study of 
	insulator-to-metal transition in Cr- and Ru-doped 
	$\rm Nd_{1/2}A_{1/2}Mn_{1-y} O_{3}$ (A=Ca, Sr)}

\author {J. -S. Kang$^1$, J. H. Kim$^1$, 
	A. Sekiyama$^2$, S. Kasai$^2$, S. Suga$^2$, 
	S. W. Han$^3$, K. H. Kim$^3$, 
	E. J. Choi$^4$, 
	T. Kimura$^5$, 
	T. Muro$^6$, Y. Saitoh$^7$, 
	C. G. Olson$^8$, 
	J. H. Shim$^9$, and B. I. Min$^9$}

\address{$^1$Department of Physics, The Catholic University of Korea,
        Puchon 420-743, Korea}

\address{$^2$Department of Material Physics, Graduate School 
	of Engineering Science, Osaka University,
        Osaka 560-8531, Japan}

\address{$^3$Department of Physics, Gyeongsang National University,
        Chinju 660-701, Korea}

\address{$^4$Department of Physics, The University of Seoul, 
	Seoul 130-743, Korea}

\address{$^5$Department of Applied Physics, University of Tokyo, 
	Tokyo 113-0033, Japan}

\address{$^6$Japan Synchrotron Radiation Research Institute (JASRI), 
	SPring-8, Hyogo 679-5198, Japan}

\address{$^7$Department of Synchrotron Radiation Research, 
	Kansai Research Establishment, 
	Japan Atomic Energy Research Institute (JAERI), SPring-8, 
	Hyogo 679-5148, Japan}

\address{$^6$Korea Research Institute of Standards and Science,
        Taejon 305-600, Korea}

\address{$^8$Ames Laboratory, Iowa State University,
        Ames, Iowa 50011, U.S.A.}

\address{$^9$Department of Physics, Pohang University of Science and
	Technology, Pohang 790-784, Korea}

\date{\today}
\maketitle

%$\rm Nd_{1/2}A_{1/2}MnO_3$ (A=Ca, Sr)
%$\rm Nd_{1/2}A_{1/2}Mn_{1-y}T_yO_3$ (A=Ca, Sr; T=Cr, Ru)
%$\rm Nd_{1/2}A_{1/2}Mn_{1-y}T_yO_{3}$ (A=Ca, Sr; T=Mn, Ru)

\begin{abstract}

Electronic structures of very dilute Cr- or Ru-doped 
Nd$_{1/2}A_{1/2}$MnO$_{3}$ (NAMO; $A$=Ca, Sr) manganites have been 
investigated using the Mn and Cr $2p \rightarrow 3d$ resonant 
photoemission spectroscopy (PES).
All the Cr- and Ru-doped NAMO systems exhibit the clear metallic
Fermi edges in the Mn $e_g$ spectra near $\rm E_F$,
consistent with their metallic ground states.
The Cr $3d$ states with $t^3_{2g}$ configuration are at $\sim 1.3$ eV 
below $\rm E_F$, and the Cr $e_{g}$ states do not participate in 
the formation of the band near $\rm E_F$.
Cr- and Ru-induced ferromagnetism and insulator-to-metal transitions 
can be understood with their measured electronic structures. 

\end{abstract} 

]

\narrowtext
%==============================================================================
%\section{Introduction}
%\label{sec:intro}
   
The Mn-site doping by magnetic cations, such as Cr, Ru, is known
to be a very efficient method to induce metallicity and ferromagnetism
in the insulating and antiferromagnetic (AFM)
Nd$_{1/2}A_{1/2}$MnO$_{3}$ (NAMO; $A$=Ca, Sr)
\cite{Rave97,Maig97,Mori99,Kimu99}.
Undoped NAMO has the charge/orbital ordered (CO/OO) insulating 
phase with the CE-type AFM spin ordering.   
By doping Ru to NAMO, the Curie temperature $\rm T_C$ is even enhanced 
\cite{Maig01}.
To explain the Cr-induced insulator-to-metal (I-M) transition,
the relaxor-ferromagnet has been proposed \cite{Kimu99},
in which the field-induced I-M transition takes place due to
the growth of ferromagnetic microclusters in a matrix of CO/OO state.
On the other hand, two valence states of Ru ions are proposed
to explain the Ru-induced ferromagnetism\cite{Martin01}.
However, these proposals have not been proved yet and the different 
behavior of Cr and Ru doping 
is not well understood.

To explore the melting mechanism of CO/OO in these systems,
the main issues to be resolved are: (i) the valence and spin states 
of impurities and (ii) the role of impurities in the metallic band 
formation.
Photoemission spectroscopy (PES) provides direct information on 
the electronic structures of the CO manganites \cite{Chain97,Seki99}. 
In this paper, we report the {\it bulk-sensitive} \cite{Kang02}
high-resolution valence-band photoemission spectroscopy (PES) study 
for Cr- and Ru-doped NAMO manganites,
including resonant photoemission spectroscopy (RPES) 
\cite{Seki99,Kang02,Suga95} near the Mn and Cr $2p$ and Nd $4d$ and 
$3d$ absorption edges.
Using the large resonance enhancement in the $2p \rightarrow 3d$ RPES 
\cite{Seki99}, we were able to measure clearly the Cr $3d$ emission 
for a very dilute Cr-doped system: 
Nd$_{1/2}A_{1/2}$Mn$_{1-y}$Cr$_y$O$_{3}$ ($A=$Ca, Sr; y=0.05, 0.07),
corresponding to only $\sim 1$ atomic $\%$.
Therefore this work opens up the possibility of directly observing 
the local electronic structure of very dilute transition-metal systems.
Furthermore, our PES measurements for a wide photon energy 
($h\nu$) range ($h\nu\approx 20$ eV $-$ 1000 eV) allow us 
to determine various partial spectral weight distributions (PSWs)
by using the cross-section effect over a wide $h\nu$ range. 
We have also made a comparison with the Ru-doped 
$\rm Nd_{1/2}Sr_{1/2}Mn_{1-y}Ru_yO_{3}$ (y=0.05).

%\section{Experimental and Calculational Details}
%\label{sec:exp}

Nd$_{1/2}A_{1/2}$Mn$_{1-y}T_{y}$O$_{3}$ single crystals  
($A$=Ca, Sr; $T$=Cr, Ru; $0 \le y \le 0.1$) were grown using 
the floating zone method. The details of the crystal growing method 
is described elsewhere \cite{Kimu99}.
High-resolution T $2p \rightarrow 3d$ RPES (T=Cr, Mn) experiments were 
performed at the twin-helical undulator beam line BL25SU of SPring-8 
equipped with a SCIENTA SES200 analyzer \cite{Saitoh00}.
Samples were fractured and measured in vacuum better than 
$\rm 3 \times 10^{-10}$ Torr at $\rm T \lesssim$ 20 K.
PES data were obtained in the transmission mode, with 
the overall energy resolution [FWHM : full width at half maximum]
of about $80$ meV at $h\nu\sim 600$ eV.
All the spectra were normalized to the photon flux estimated
from the mirror current.  
Low energy PES experiments were carried out at the Ames/Montana 
beamline at the Synchrotron Radiation Center (SRC).
Samples were fractured and measured in vacuum with a base pressure 
better than $\rm 3 \times 10^{-11}$ Torr and at $\rm T \lesssim$ 15 K
with the FWHM $\approx 80$ meV at $h\nu\approx 20$ eV.
The fractured surfaces were rough and the measured spectra showed 
no angle-dependence, suggesting that our PES data could be considered 
as being angle-integrated.
All the samples showed a clean single peak in the O $1s$ core-level 
spectra and no 9 eV binding energy (BE) peak in the valence-band
spectra.

%\section{Results and Discussion}
%\label{sec:results}

\begin{figure}
\epsfig{file=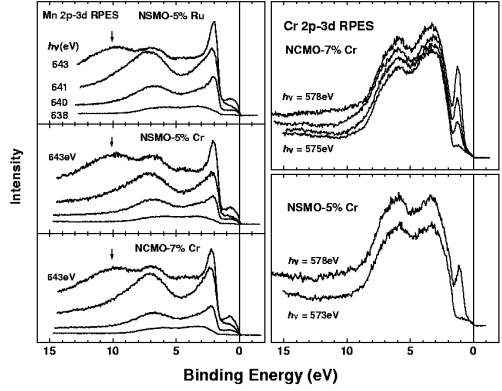,angle=270,width=8.65cm}
\caption{Left: The Mn $2p \rightarrow 3d$ RPES spectra 
	of NSMO-5$\%$Ru, NSMO-5$\%$Cr, and NCMO-7$\%$Cr
	around $h\nu \sim 640$ eV. 
	The bump structures below the arrow are due to the Mn LMM Auger 
	emission.
	Right: The Cr $2p \rightarrow 3d$ RPES spectra of NCMO-7$\%$Cr 
	and NSMO-5$\%$Cr around $h\nu \sim 575$ eV. }
\label{rpes}
\end{figure}

Figure~\ref{rpes} shows the valence-band spectra of Cr- and Ru-doped 
$\rm Nd_{1/2}Ca_{1/2}MnO_3$ (NCMO) and $\rm Nd_{1/2}Sr_{1/2}MnO_3$
(NSMO) near the Mn and Cr $2p_{3/2}$ absorption edges.
Large enhancement is observed across the Mn ($ h\nu =643$ eV) and Cr
($ h\nu =578$ eV) absorption edges.
Our Mn $2p \rightarrow 3d$ RPES spectra for Cr- and Ru-doped NCMO 
and NSMO are similar to those of the previous report on CO NSMO 
\cite{Seki99}. 
As to the resonating behavior of the Mn $3d$-derived states (the left 
panel), those near $\sim 2.3$ eV BE and $\rm E_F$ ($0 \sim 1$ eV) 
are identified as the $t^3_{2g}$ and $e^{x}_{g}$ ($x \approx 0.5$)
majority-spin states, respectively.  
These RPES measurements reveal that the high BE features 
($5 \sim 8$ eV) also have the large Mn $3d$ electron 
character, which is strongly hybridized with the O $2p$ electrons. 
The broad bump around 10 eV BE (denoted by arrow) for $h\nu\sim 643$ eV
is due to the Mn LMM Auger emission. 

\begin{figure}
\epsfig{file=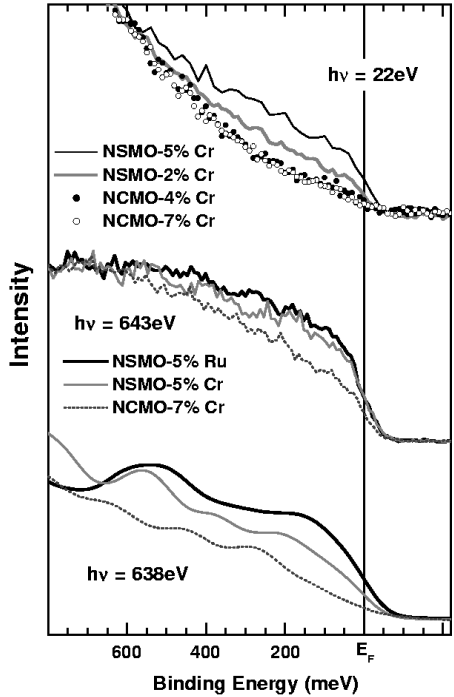,width=7.65cm}
\caption{Near-$\rm E_F$ region of the valence-band spectra of 
	$\rm Nd_{1/2}A_{1/2}Mn_{1-y}T_{y}O_{3}$ (A=Ca, Sr; T=Cr, Ru)
	at $h\nu=22$ eV (top) and at the Mn $2p \rightarrow 3d$
	on-resonance ($h\nu\approx 643$ eV) (middle).
	These spectra were obtained with $\sim 80$ meV at FWHM.
	Bottom: the Near-$\rm E_F$ spectra at the Mn $2p \rightarrow 3d$
	off-resonance ($h\nu\approx 638$ eV), obtained with $\sim 150$
	meV at FWHM.  }
\label{ef}
\end{figure}

In the right panel, the sharp resonating feature around $\sim 1.3$ eV 
BE indicates strong Cr $3d$ character. 
The absence of the Cr $3d$ electron emission near $\rm E_F$
indicates that doped Cr ions in NCMO and NSMO are in the localized 
trivalent Cr$^{3+}$ states ($t^3_{2g}$ configuration), 
and that Cr $e_{g}$ states are located above the Mn $e_{g}$ states. 
In fact, our Cr $2p$
XAS spectrum of NAMO (not shown here) is very similar to that of
a formally trivalent Cr$_2$O$_3$ \cite{Theil99}, but quite different
from those of formally divalent or tetravalent Cr compounds.
It thus suggests that Cr $e_{g}$ states do not 
participate in the formation of the band near $\rm E_F$, 
nor affect the bandwidth of the $e_g$ states near $\rm E_F$.
This feature can be understood based on the previous finding that 
Cr spins order antiparallel to the Mn subnetwork \cite{Toul99},
and so the hopping between Cr and neighboring Mn ions is not allowed.
This conclusion is supported by the comparison of the $e_g$ spectra 
near $\rm E_F$ in Fig.~\ref{ef}. 

\begin{figure}
\epsfig{file=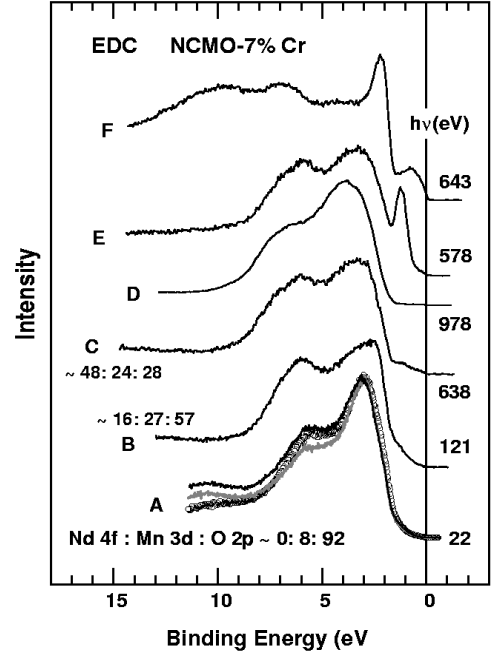,width=7.65cm}
\caption{Valence-band spectra of NCMO-7$\%$Cr for a wide $h\nu$ 
	range (22 eV $\le h\nu \le$ 978 eV).
	The $h\nu=978$ eV spectrum is that for NSMO-5$\%$ Ru,
	which is essentially the same as for NCMO-7$\%$Cr.
	See Ref.\protect\cite{978} for the details.} 
\label{edc}
\end{figure}

Figure~\ref{edc} compares the valence-band spectra of 7$\%$ Cr-doped 
NCMO for a wide $h\nu$ range (22 eV $\le h\nu \le 978$ eV).
The different valence-band line shapes with varying $h\nu$
reflect different contributions from different electron character. 
Note that these high-$h\nu$ PES spectra can be considered to represent 
the bulk electronic states \cite{Kang02,Seki00}.  
The top three spectra represent the on-resonance spectra
in the Mn $2p \rightarrow 3d$, Cr $2p \rightarrow 3d$, and 
Nd $3d \rightarrow 4f$ RPES.
Therefore the enhanced features in the top three spectra are due to 
the Mn $3d$, Cr $3d$, and Nd $4f$ electron emissions.
The resonance enhancement for the Nd $3d \rightarrow 4f$ RPES is so 
large that the on-resonance spectrum ($h\nu=978$ eV) can be considered 
as the Nd $4f$ partial spectral weight (PSW) 
\cite{978}. 

If one assumes Nd$^{3+}$ ($4f^3$), Mn$^{3.5+}$ ($3d^3-3d^4$), and the 
filled O $2p$ bands ($2p^6$) in NCMO, and ignores the resonance effect,
then the cross-section ratio of Nd $4f$ : Mn/Cr $3d$ : O $2p$ electrons
per unit cell is about 
$\sim 1 \%$ : $\sim 8 \%$ : $\sim 92 \%$ at $h\nu\approx 22$ eV, 
$\sim 16 \%$ : $\sim 27 \%$ : $\sim 57 \%$ at $h\nu\approx 120$ eV,
and $\sim 48 \%$ : $\sim 24 \%$ : $\sim 28 \%$ at $h\nu\approx 640$ eV
\cite{Yeh85}. 
Therefore the $h\nu=22$ eV spectrum can be considered as the O $2p$ PSW.
In the the $h\nu=22$ eV spectra, open circles and gray lines denote 
that for 2$\%$ Cr-doped NSMO and 4$\%$ Cr-doped NCMO, respectively.
The similar lineshapes at $h\nu=22$ eV indicate that the O $2p$
states are very similar in Cr-doped NAMO manganites.
As $h\nu$ increases, the Mn $3d$ and Nd $4f$ emissions increase with
respect to the O $2p$ emission, and the Mn $3d$ and O $2p$ emissions 
become comparable to each other at $h\nu\approx 638$ eV 
(the Mn $2p \rightarrow 3d$ off-resonance).
In fact, we have observed that the overall features of Cr- and Ru-doped
NAMO (A=Ca, Sr) are very similar at $h\nu=638$ eV, and that 
the valence-band widths for NCMO systems are slightly 
narrower than those for NSMO systems \cite{Unpub}, 
which is consistent with the larger ionic size of Sr than for Ca, 
causing the electron hopping easier.

The Nd $4f$-derived states exhibit two broad peaks around $\sim 4$ eV 
and $\sim 7$ eV BE, which overlap substantially with the O $2p$ states 
between $3-6$ eV (see the $h\nu=22$ eV spectrum). 
No emission is observed near $\rm E_F$ where the Mn $e_g$ bands are 
located. 
These observations suggest that Nd $4f$ states are strongly hybridized 
with the O $2p$ states, but very little with the Mn $e_g$ states. 
Thus the Nd $4f$ states do not participate in the band formation 
near $\rm E_F$. 
We have found \cite{Unpub} that the Nd $4f$ PSWs of Cr- and Ru-doped 
NAMO (A=Ca, Sr), determined from the Nd $4d \rightarrow 4f$ RPES, are 
essentially the same as one another and also very similar to that 
determined from the Nd $3d \rightarrow 4f$ RPES.  
Considering the difference in the electron mean free paths between 
the Nd $4d \rightarrow 4f$ RPES and the Nd $3d \rightarrow 4f$ RPES, 
this finding indicates that the {\it surface} Nd $4f$ states are very 
similar to the {\it bulk} Nd $4f$ states in NAMO manganites, which 
makes a sharp contrast to the case of Ce compounds \cite{Seki00}.
We interpret the Nd $4f$ PSW to represent roughly the $4f^3$ ground 
state, but with the large final-state hybridization to O $2p$ orbitals
\cite{Suga95,Kang99}.

Figure~\ref{ef} compares the near-$\rm E_F$ regions of the valence-band 
spectra of Cr- or Ru-doped NCMO and NSMO at $h\nu=22$ eV (top), 
and at the on-resonance energy of the Mn $2p \rightarrow 3d$ RPES 
($h\nu\approx 643$ eV) (middle). 
The $h\nu\approx 643$ eV spectra (middle) correspond to the 
Mn $e_g$ PSWs. 
At $h\nu\approx 643$ eV, the metallic Fermi edges are clearly observed 
for all the samples, which is consistent with their metallic ground 
states. In contrast, the spectral intensity near $\rm E_F$ 
($\rm I(E_F)$) in the $h\nu=22$ eV is very weak in view of the metallic
nature of the samples. 
This difference arises from the fact the O $2p$ emissions are dominant
at $h\nu=22$ eV and that the O $p$ states have a negligible 
contribution near $\rm E_F$. 
This comparison reveals that Mn $e_g$ states play an important role
in determining the I-M transition.
For comparison, the spectra at the off-resonance energy of the Mn $2p 
\rightarrow 3d$ RPES ($h\nu\approx 638$ eV) are shown at the bottom, 
which were obtained with a lower resolution (FWHM $\sim 130$ meV) 
than the above spectra (FWHM $\sim 80$ meV) \cite{GB}.   
For all the cases, $\rm I(E_F)$ is lower for the NCMO-based system 
than for the NSMO-based system, reflecting the stronger metallic nature
for NSMO than for NCMO. 
Note that the larger Cr-doped NSMO exhibits a higher $\rm I(E_F)$ 
than the smaller Cr-doped NSMO (see $h\nu=22$ eV), and that 
the Ru-doped NSMO shows a higher $\rm I(E_F)$ than the Cr-doped NSMO 
(see $h\nu=638$ eV). 
These features are consistent with the observations that the metallic
nature increases with the higher Cr-doping and that $\rm T_C$ is 
higher for the Ru-doped NSMO than for the Cr-doped NSMO.

\begin{figure}
\epsfig{file=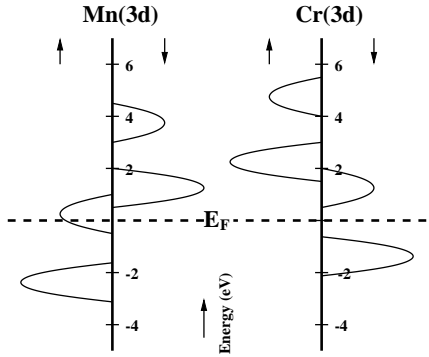,angle=270,width=7.65cm}
\caption{Schematic local density of states of Mn and Cr 
	for Cr-doped NAMO.  }
\label{diagram}
\end{figure}

Now let us discuss the melting mechanism in Cr- and Ru-doped NAMO,
based on our observations. The undoped CO/OO NAMO is expected to have 
the ordered Mn$^{3+}$/Mn$^{4+}$ ($3d^4/3d^3$) configuration 
with the CE-type AFM spin ordering.   
We have found that Cr $t_{2g}$ states are located well below $\rm E_F$,
resulting in the trivalent Cr$^{3+}$ state ($d^3$), 
as shown in the schematic diagram for the local density of states  
(Fig.~\ref{diagram}). 
Then the substitution of a Cr$^{3+}$ ($d^3$) for a Mn$^{3+}$ ($d^4$) 
site in the CO/OO NAMO will correspond to the hole filling 
in the undoped NAMO, and it will play a role similar 
to the substitution of a Mn$^{4+}$ ion for a Mn$^{3+}$ ion.
So one can simply conjecture that the CO/OO insulating phase is 
transformed into the metallic phase by breaking the commensurate 
quarter filling state. For example, the phase diagram of NSMO manifests
that, with increasing the hole concentration, the CO/OO insulating 
phase of half-doped NSMO with the CE-type AFM is transformed into 
the metallic A-type AFM \cite{Brink99,Min02}.
Similarly, one can also infer that the substitution of Ru$^{4+}$ 
($d^4$) into a Mn$^{4+}$ ($d^3$) site will correspond to the electron 
filling, and so the CO/OO insulating phase of half-doped NSMO is 
transformed into the metallic ferromagnetic phase.

The above picture of the filling control, however, is not complete
since Cr-doped NSMO is not A-type AFM but ferromagnetic.
Moreover, the stronger CO system NCMO would not exhibit the metallic 
phase merely with the filling control.
It is thus likely that, not just the valence state but also
other characteristics such as the spin state and the
electronic structure are important.
The localized spin of Cr$^{3+}$ affects the spin ordering.
The AFM superexchange interaction between Cr$^{3+}$ and neighboring 
Mn$^{4+}$ ions will give rise to the so-called {\it domino effect} 
of reversing the spin directions of Mn$^{3+}$/Mn$^{4+}$ in the 
ferromagnetic zig-zag chain of the CE-type AFM \cite{Martin01}.
Likewise, the FM superexchange interaction between Ru$^{4+}$ and 
neighboring Mn$^{3+}$ ions also gives rise to the domino effect of
reversing the spin directions of Mn$^{3+}$/Mn$^{4+}$.
In addition, since Cr$^{3+}$ is not a Jahn-Teller active ion,
it will play as a defect for the OO and
the hopping strength will be enhanced around Cr ions due to the
reduced Jahn-Teller polaron narrowing effect.
Hence, the metallic phase could be formed even in NCMO at least around 
Cr ions, albeit not in whole crystal, if the double-exchange (DE) 
hopping strength becomes larger than the AFM superexchange interaction 
\cite{Brink99,Min02} or the hopping strength is larger than
the intersite Coulomb interaction between carriers \cite{Lee97}.
This picture is consistent with the concept of relaxor-ferromagnet  
\cite{Kimu99}.
Note that, in the metallic phase, Cr$^{3+}$ itself does not participate 
in the DE mechanism because two configurations of Mn$^{3+}$-Cr$^{3+}$ 
and Mn$^{4+}$-Cr$^{2+}$ are not degenerate, as observed in the PES 
above. That is, doped Cr$^{3+}$ would just play a role of initiating 
the hopping of carriers in the system.

%\section{Conclusions}
%\label{sec:conc}

To summarize,
the electronic structures of very dilute Cr- and Ru-doped NAMO
have been investigated by employing RPES. 
The large resonance enhancement in the $2p \rightarrow 3d$ RPES
allows us to observe clearly the Cr $3d$ emission in very 
dilute ($\sim 1$ atomic $\%$) Cr-doped NAMO. 
The Cr $3d$ states are observed at $\sim 1.3$ eV below $\rm E_F$,
corresponding to the $t^3_{2g}$ configuration of Cr$^{3+}$ ions. 
This finding suggests that Cr $e_{g}$ states are located above the Mn 
$e_{g}$ states and they do not participate in the formation of the band 
near $\rm E_F$. All the Cr- and Ru-doped NAMO systems exhibit the clear
metallic Fermi edges in the Mn $e_g$ spectra near $\rm E_F$,
consistent with their metallic ground states.
The spectral intensity at $\rm E_F$
is higher for the NSMO-based system than for the NCMO-based system,
reflecting the stronger metallic nature for NSMO than for NCMO because
of the larger ionic size of Sr than for Ca.
Further, Ru-doped NSMO has higher spectral intensity near $\rm E_F$
than Cr-doped NSMO, consistent with a higher $\rm T_C$ for Ru-doped 
NSMO.

Acknowledgments$-$
This work was supported by the KRF (KRF--2002-070-C00038) and the KOSEF 
through the CSCMR at SNU and the eSSC at POSTECH.
PES experiments were performed at the SPring-8
(JASRI: 2001B0028-NS-np) and at the SRC (NSF: DMR-0084402).


\begin{references}
\bibitem{978} The $h\nu=978$ eV spectrum in Fig.~\ref{edc} is 
	that of NSMO-5$\%$ Ru because we do not have that spectrum 
	for NCMO-7$\%$ Cr. However, we have found  \cite{Unpub} that 
	the Nd $4f$ PSW is essentially the same for Cr- and Ru-doped 
	NCMO and NSMO.
\bibitem{GB} The off-resonance spectra shown at the bottom have 
	been smoothened by using a Gaussian function of 100 meV at FWHM
	to reduce the signal-to-noise ratio. However, the trend
	observed in the raw spectra is still maintained.
\end{references}
\end{document}